**Methane and Oxygen from Energy-efficient, Low Temperature *in-situ* Resource Utilization (ISRU) Enables Missions to Mars**

Mohamed Shahid [a,d,1], Bradley Chambers [a,b,1], and Shrihari Sankarasubramanian [a,b,c,*]

[a] Department of Biomedical Engineering and Chemical Engineering, University of Texas at San Antonio, San Antonio, TX 78249, USA.

[b] NASA MIRO Center for Advanced Measurements in Extreme Environments (CAMEE), University of Texas at San Antonio, San Antonio, TX 78249, USA.

[c] Texas Sustainability Research Institute (TSERI), University of Texas at San Antonio, San Antonio, TX 78249, USA.

[d] Department of Chemical Engineering, University of Petroleum and Energy Studies, Dehradun 248007, India

[1] These authors contributed equally.

[*]Corresponding author e-mail: Shrihari.sankarasubramanian@utsa.edu

## Abstract

NASA's mandate is a human mission to Mars in the 2030s and sustained exploration of Mars requires *in-situ* resource utilization (ISRU). Exploiting the Martian water cycle (alongside perchlorate salts that depress water's freezing point to < 213K) and the available 95 vol.% atmospheric $CO_2$, we detail an ultra-low temperature (255K) $CO_2$-$H_2O$ electrolyzer to produce methane fuel and life-supporting oxygen on Mars. Our polarization model fit experimental Martian brine electrolyzer performance and predicted $CO_2$ electrolysis occurring at comparatively lower potentials (vs. water electrolysis) on Mars. A hypothetical 10-cell, 100cm$^2$ electrode-area-per-cell electrolyzer produced 0.45$gW^{-1}day^{-1}$ of $CH_4$ and 3.55$gW^{-1}day^{-1}$ of $O_2$ at 2V/cell and 50% electrolyzer faradaic efficiency vs. a best-case production of 2.5$gW^{-1}day^{-1}$ of $O_2$ by the Mars Oxygen *in-situ* Resource Utilization Experiment (MOXIE) from NASA's Mars 2020 mission (MOXIE produces no fuel). Material performance requirements are presented to advance this technology as an energy-efficient complement to MOXIE.

**Plain Language Summary**

Future sustained human missions to Mars will require astronauts to "live off the land" through *in-situ* Resource Utilization (ISRU). This is necessitated by the immense energetic cost of moving material out of Earth's gravity well – completely provisioning a Mars mission from Earth, including 35 metric tons of propellant needed for the return journey, is estimated to require *ca* 400 metric tons of propellant (fuel and oxidant) on 4-5 heavy lift launch vehicles. As a cost-effective alternative, NASA is considering deploying a high temperature solid oxide electrolyzer 26 months in advance of a human mission to produce 25-30 metric tons of oxidant ($O_2$) using atmospheric $CO_2$ abundantly present on Mars (based on technology demonstrated through the Mars Oxygen *in-situ* Resource Utilization Experiment (MOXIE) from NASA's Mars 2020 mission). Remedying shortcomings of this system, we propose an integrated ultra-low temperature electrolyzer that produces both fuel and oxidant utilizing Martian atmospheric $CO_2$ in conjunction with liquid brines present on Mars. Thus, our system produces the same oxidant as MOXIE with the added benefit of producing the other propellant component, methane ($CH_4$), all at lower energy consumption. Given the geographic limitations on brine availability, we envision our system being an able complement to MOXIE to advance the goal of sustained human exploration of Mars.

## Introduction

The United States National Aeronautics and Space Administration (NASA)'s current goal is to land humans on Mars in the 2030's. The exploration of Mars is also a priority for several national and private space entities with lander missions from the US and China and orbiters from the US, China, Europe, Russia, India, and the United Arab Emirates (UAE) presently active there[1]. The duration of any mission to Mars is circumscribed by the constraints imposed by the mass that can be sent to Mars from Earth. For example, SpaceX's Falcon Heavy (weighing 1420 metric-tons fully loaded) is designed to deliver a 16.8 metric-ton payload to Mars (1.8% of total weight). The immense energetic cost of moving material out of Earth's gravity well is illustrated by the following example - completely provisioning a Mars mission from Earth, including 35 metric tons of propellant needed for the return journey, is estimated to require *ca* 400 metric tons of propellant (fuel and oxidant) on 4-5 heavy lift launch vehicles[2]. Additionally, astronauts consume 0.8 to 1.2kg of oxygen a day depending on activity, sex, height, and weight[3]. Thus, any economical long-term (weeks to months) mission necessitates the exploitation of resources present on Mars for life-support and energy[4].

Key inputs for *in-situ* resource utilization (ISRU) have been identified in the Martian atmosphere and surface. The Martian atmosphere significantly differs from that of Earth's with its predominant constituent being $CO_2$ and the atmospheric pressure on Mars (636 Pa) is significantly lower than that of Earth (101325 Pa) (detailed in **Table 1**). Critically, the relatively low diurnal temperature range on Mars also suggests a heating energy penalty for any ISRU process carried out there.

In addition to abundant $CO_2$ in the Martian atmosphere, the comprehensive elucidation of Martian regolithic geochemistry by a succession of robotic landers and orbiter missions has led to the identification of deliquescent perchlorate salts as outlined in our group's previous work[5]. This includes the notable discovery of significant quantities of perchlorate and sulfate salts of sodium and magnesium by NASA's Phoenix lander (through its wet chemistry instrument (WCI[6])) and optically observed sublimation of water ice discovered under a few inches of regolith[7]. Multiple pathways allow for the existence of water on (or under) the surface of present-day Mars through premelting[8], by adsorption of atmospheric vapor at grain-ice boundaries[5] and by greenhouse melting creating three phases of water[9], with one phase being a temporary liquid water at the top of the subsurface ice[10]. Perchlorate salts play an important role in any extant hydrological cycle on Mars by depressing the freezing point of water to *ca* 203K as shown in our group's previous work[5], with Martian water phase thermodynamic analysis[11] and with solubility polytherms[12]. This freezing point depression allows for the existence of liquid water, a key ISRU feedstock, which is further corroborated by the phoenix lander site's evidence for liquid water[13] and saline water stability on Mars[14]. NASA's Spirit and Opportunity rovers found historic evidence of super-cold local acidic brines through observation of acidic aqueous activity, evaporation, and desiccation of the Martian regolith[15]. Contemporary liquid water flows are also indicated by the geological observations of the Mars Reconnaissance Orbiter (MRO[16]). Furthermore, $2x10^{18}$ kg of water on Mars is present in the form of polar ice caps[17] as evidenced by the Mars Odyssey Gamma Ray Spectrometer (GRS[18]) and Mars Express spacecrafts. The Mars Advanced Radar for Subsurface and Ionosphere Sounding (MARSIS) instrument on-board the Mars Express spacecraft has also detected multiple sub-glacial water bodies underneath the Martian south pole at Ultimi Scopuli[19]. Thus, abundantly present $CO_2$ and $H_2O$ have been chosen as ISRU feedstock.

**State-of-the-art and Proposed New Direction in ISRU**

NASA's Mars Oxygen *in-situ* Resource Utilization Experiment (MOXIE) on the Perseverance rover is the most advanced Martian ISRU demonstration till date. This high temperature (1073K) solid oxide electrolyzer testbed produces $O_2$ and CO from atmospheric $CO_2$ through the reaction shown below[20]:

$$2CO_2(g) = 2CO(g) + O_2(g) \quad (1)$$

Presently, the 10-cell solid oxide electrolyzer (SOXE) stack (2x 5-cell stacks in series with electrode area = 22.7 $cm^2$) in MOXIE has been reported to operate at 50 % Faradaic efficiency (F.E) with a total power requirement of ~300 W[21]. Of the 300W, *ca* 115 W is expended operating the stack with the rest powering the balance of plant. The stack operation in turn consumes *ca* 35 W[21] (8.7 V stack operating voltage and maximum operating current of 4 A) for carrying out the electrolysis and *ca* 80 W for maintaining the stack temperature at 1037 K.

MOXIE has been reported to produce a maximum of 12 g/ hr of $O_2$[22] which translates to a power normalized $O_2$ production of 8.27 grams.$W^{-1}$.$day^{-1}$ (considering only the power required for electrolysis, *ca* 35 W). But considering the total power requirements to run the stack (the SOXE will not run unless it is heated to or near its operating temperature of 1037 K)[21] and normalizing the $O_2$ production rate using this value (*ca* 115W), the power normalized $O_2$ production rate was 2.5 grams.$W^{-1}$.$day^{-1}$ under MOXIE operating conditions on Mars. We will show that our electrolyzer modeled using a 10-cell stack (electrode active area = 100 $cm^2$) can provide a higher (~1.4x) $O_2$ production operating at the same 50 % electrolyzer F.E.

ISRU systems could potentially be sent to the surface 26 months before a manned mission in order to produce the 30 metric tons of oxygen needed to support a human mission to Mars. Proposals exist for a larger 1 metric ton MOXIE-like electrolyzer to produce 25-31 metric tons of oxygen for life support and oxidant[23]. Despite MOXIE's successful production of $O_2$, the need for downstream processing to remove CO and the inherent risks of CO toxicity implies increased balance of plant (BOP) requirements and reduced overall energy efficiency. Furthermore, high operational temperatures introduce issues of thermal waste and, critically, safety. By running our system at (or close to) the Martian ambient temperature and producing $O_2$ from $H_2O$ (and $CH_4$ from $CO_2$), we propose to avoid both these potential safety risks inherent in the MOXIE system.

Another critical design consideration is the nature of engines powering the descent stage landing on Mars. Methalox engines, utilizing cryogenic methane as fuel and cryogenic oxygen as oxidant, are the preferred power source (a listing, representative of global efforts towards developing methalox engines, is presented in **Table 2**). An issue for future missions is that MOXIE only produces one component used in methalox engines, $O_2$, leaving the need for the other propellant component, $CH_4$, unfulfilled. State-of-the-art methalox rocket engines require 12 to 13 metric tons of propellant (oxygen and methane) to transfer a metric ton of payload from low earth orbit[24]. The economic value of the 35 metric tons of propellant needed for a return journey from Mars can be estimated from the Perseverance rover's cost of ~$237,000 per kg of payload to the Martian Surface[25]. Using this proven cost gives a potential savings of ~$8.3 billion if propellant is created on the Martian surface, with ~$1.66 billion of that cost being related to sending the methane needed for the methalox rocket engine propellant. The key aim of our electrolyzer is to model the production of both, methane (for methalox rockets) and oxygen (for life support and oxidizer) using ISRU of $CO_2$ at Martian conditions.

Sankarasubramanian and co-workers have previously demonstrated that electrolysis of Martian brines can be successfully carried out at the average Martian ambient temperature (237K) to produce $H_2$ fuel and life-support $O_2$ with significantly improved energy efficiencies compared to MOXIE[5]. Building on this work, the ultra-low temperature $CO_2$-brine electrolyzer (**Fig. 1**) can produce methane as fuel (in addition to $O_2$) through the following full cell reaction:

$$CO_2(g) + 2H_2O(l) = CH_4(g) + 2O_2(g) \quad (2)$$

We initially anticipated electrocatalysis being an impediment to realizing this electrolyzer. The average Martian diurnal temperature ranges between 184K to 242K. Thus, based on Arrhenius kinetics, it was anticipated that the rate of 8-electron $CO_2$ reduction to $CH_4$ would be significantly slower than at terrestrial conditions. Fortuitously, recent studies have demonstrated that Cu electrocatalysts exhibit anti-Arrhenius kinetics and, counterintuitively, improved $CH_4$ selectivity going from 293K to 255K[26]. This is attributed to the solubility of $CO_2$ increasing and the $H_2$-evolution reaction (HER) kinetics decreasing as temperature decreases. At these lower temperatures, it was demonstrated that gas hydrate crystalline structures (clathrates) restrict rotation and translation of gas molecules[27] and influence carbon dioxide reduction reaction ($CO_2$RR) selectivity. Furthermore, at these lower temperatures, the competing HER pathway is shifted towards negative overpotentials, and is directly inhibited by adsorbed CO from the $CO_2$RR, leading to much higher selectivity and faradaic efficiency for $CO_2$ reduction on Cu electrodes[28]. We anticipate the $CH_4$ and (unreacted) $CO_2$ exiting the electrolyzer being separated using cryogenic distillation and the lower Martian ambient temperature enabling down-stream cryogenic distillation (**Figure 1**) with a lower cooling energy penalty. However, the general electrolyzer design is the primary focus of this polarization model. We have already demonstrated $O_2$ production at acceptable rates and selectivity under Martian conditions in our previous work[5]. Ultimately, some of the conditions that make Mars uninhabitable for humans are anticipated to be beneficial for the synthesis of methalox engine propellant using a liquid brine and $CO_2$ electrolysis system.

We evaluated the operational viability and materials requirements for the $CO_2$-$H_2O$ electrolyzer by building thermodynamic and electrochemical polarization models of this system. The models were first validated against our experimental data for the operation of a Martian $H_2$-$O_2$ electrolyzer. For the proposed $CO_2$-$H_2O$ electrolyzer, the Nernst potential, the choice of anode/cathode and overpotentials for the electrocatalyst at the anode/cathode were correspondingly calculated or parameterized from experiments in the literature to model the performance of these electrolyzers under Martian conditions. Finally, we modeled the fuel and oxidant generation capacity of a conservatively sized electrolyzer stack consisting of 10-cells of 100cm$^2$ active area each and show that this system is an energy-efficient method to provision long-term missions to Mars.

**Electrolyzer operation on Mars**

Thermodynamic and polarization models were built for the reactions depicted below. Brine electrolysis (**eqn (3)**) to produce $H_2$ and $O_2$ served as model validation against existing experimental data[6].

$$H_2O(l) = H_2(g) + \frac{1}{2}O_2(g) \qquad \textit{(Brine − electrolysis) (3)}$$

$$CO_2(g) + 2H_2O(l) = CH_4(g) + 2O_2(g) \qquad (CO_2 - \textit{Brine electrolysis) (4)}$$

The corresponding half-cell reactions (and the standard electromotive force (emf) for the reaction ($E^0_{rxn}$)) for brine electrolysis to $H_2$ and $O_2$ are –

$$H_2O = \frac{1}{2}O_2 + 2H^+ + 2e^- \ (E^0_{rxn} = -1.23V) \qquad (5)$$

$$2H^+ + 2e^- = H_2 \ (E^0_{rxn} = 0V) \qquad (6)$$

And the corresponding half-cell reactions for $CO_2$-brine electrolysis to $CH_4$ and $O_2$ are [29]–

$$4H_2O = 2O_2 + 8H^+ + 8e^- \; (E^0_{rxn} = -1.23V) \tag{7}$$

$$CO_2 + 8H^+ + 8e^- = CH_4 + 2H_2O \; (E^0_{rxn} = -0.17V) \tag{8}$$

**Thermodynamic model development**

The thermodynamic property change (*dS*, *dH* and *dG*) over the course of the reactions depicted in **eqn (3)** and **eqn (4)** were obtained at Martian conditions. i.e., between 230K to 255K. The change in enthalpy ($dH$) was calculated using the following relationship –

$$dH = H_{298K} + c_p \times (T - T_{ref}) \tag{9}$$

The change in entropy ($dS$) was calculated using the following relation with the specific heat capacity ($c_p$) –

$$dS = S_{298K} + c_p \times ln\left(\frac{T}{T_{ref}}\right) \tag{10}$$

Having obtained $dH$ and $dS$ for the given reaction, the free energy change (*dG*) was calculated as shown below-

$$dG = dH - T\,dS \tag{11}$$

*dG* was in turn used to calculate the reversible electromotive force ($E_{emf}$) i.e., the minimum thermodynamic potential required for the reactions to occur, using the following equation -

$$\Delta G = -nF|E_{emf}| \tag{12}$$

where the number of electrons ($n$) is either 2 or 8 depending on whether brine electrolysis or $CO_2$-brine electrolysis is carried out by the electrolyzer, and F is Faraday's constant (96,485 C mol$^{-1}$).

The $c_p$ of supercooled water has been measured using various techniques like, differential scanning calorimetry[30], drift calorimetry[31] and proton magnetic resonance chemical shift measurements[32]. These observations consistently point to an anomalous increase in $c_p$ in inverse correlation to temperature.

For the case of brine electrolysis (**eqn (3)**), the thermodynamic properties (*dS* and *dH*) were calculated from the specific heat capacity ($\mathrm{C_p}$, J mol$^{-1}$ K$^{-1}$) of supercooled water measured using an adiabatic calorimeter[33], between 235K to 285 K, with $\mathrm{T_{ref}}$ at 298K. The data from the literature was shown to accurately fit the following empirical expression[36] -

$$C_p^{H_2O} = 0.44 \times \left(\frac{T - 222}{222}\right)^{-2.5} + 74.3 \tag{13}$$

The values of $dH$, $dS$ and $dG$ for $H_2O$ (l) can be found in **Table S1** in the Supplementary Materials. Having obtained the thermodynamic values as detailed above, we calculated $|E_{emf}| and\ E_{thermoneutral}$ across the temperature range of interest (230K to 500K) for brine electrolysis. The $|E_{emf}| and\ E_{thermoneutral}$ can be found in **Table S2** in the Supplementary Materials. We also obtained the following general empirical correlation between $|E_{emf}|$ vs. temperature ($T$) between 230K to 1300K from these calculated values -

$$|E_{emf}| = (2e^{-7} \times T^2) - (6e^{-4} \times T) + 1.43 \tag{14}$$

The thermoneutral voltage ($E_{thermoneutral}$) was calculated from the corresponding *dH* values using **eqn. (15)**.

$$E_{thermoneutral} = \frac{dH}{2F} \quad (15)$$

For the case of $CO_2$-brine electrolysis, in addition to the thermodynamic properties of $H_2O$ in liquid phase, the gas phase thermodynamic properties of $CO_2$ and $CH_4$ were required between 230K to 270K. The specific heat capacity ($C_p$, J mol$^{-1}$ K$^{-1}$) values for $CO_2$ and $CH_4$ were obtained from literature[34,35], respectively. The $dH$, $dS$ and $dG$ for $CO_2$ (g) and $CH_4$ (g) can be found in **Table S3** and **Table S4** in the Supplementary Materials. The thermodynamic properties (*dS* and *dH*) were calculated from $C_p$ for temperatures between 220-300K and 120-500K with $T_{ref}$ at 298 K. We also obtained the following empirical correlations for $c_p$ vs. *T* from this data -

$$C_p^{CO_2} = 0.039 \times T + 25.74 \quad (16)$$

$$C_p^{CH_4} = 0.0001\ \times T^2 - 0.038 \times T + 36.84 \quad (17)$$

Similar to the case of the brine electrolyzer, we obtained an empirical expression correlating $|E_{emf}|$ vs. temperature (T) between 230K to 1300K using these $dH$ and $dS$ values -

$$|E_{emf}| = (1e^{-7} \times T^2\ ) - (2e^{-4} \times T) + 1.12 \quad (18)$$

The thermoneutral voltage ($E_{thermoneutral}$) was calculated from the corresponding *dH* values using **eqn. (19)**.

$$E_{thermoneutral} = \frac{dH}{8F} \quad (19)$$

The $|E_{emf}|and\ \ E_{thermoneutral}$ can be found in **Table S5** in the Supplementary Material. These thermodynamic properties (d$G$, $dH$, $E_{emf}$ and $E_{thermoneutral}$) for brine electrolysis and $CO_2$-brine electrolysis from 230K to 500K are depicted in **Fig 2**. As a consequence of the anomalous $c_p$values, we saw that the $dG$ and hence $|E_{emf}|$ and $E_{thermoneutral}$ increased in inverse proportion to the temperature. The effect of these anomalous $c_p$values at <273K was found to be more significant than any changes due to phase transitions.

**Nernst Potentials for Electrolysis**

The Nernst equation provides information about the equilibrium potential of the electrolyzer under open circuit conditions. The equilibrium potential is the sum of contributions from reversible electromotive force (minimum thermodynamic potential) and concentration (activity) of electroactive species in a reaction.

For example, in case of brine electrolyzers operating at terrestrial atmospheric pressure of 101325 Pa, the $E_{Nernst}$ can be defined as,

$$E_{Nernst} = |E_{emf}| - \frac{RT}{2F} ln\left(\frac{[y_{H_2}][y_{O_2}]^{\frac{1}{2}}}{[y_{H_2O}]}\right) \tag{20}$$

where the mole fractions of $H_2$, $O_2$ and $H_2O$ are written as $y_{H_2}$ , $y_{O_2}$ and $y_{H_2O}$. Since the $y_{H_2O}$ is $\sim$1 in liquid phase, the expression reduced to,

$$E_{Nernst} = |E_{emf}| - \frac{RT}{2F} ln\left([y_{H_2}][y_{O_2}]^{\frac{1}{2}}\right) \tag{21}$$

We also corrected the Nernst potentials for brine electrolysis for Martian surface pressure (636 Pa) using the additional term, $\frac{P_{Mars}}{P_{Std}}$,, as shown below,

$$E_{Nernst} = |E_{emf}| - \frac{RT}{2F} ln\left( [y_{H_2}][y_{O_2}]^{\frac{1}{2}} \times \left[\frac{P_{Mars}}{P_{Std}}\right]^{\frac{3}{2}} \right) \tag{22}$$

Similarly, the Nernst potentials for $CO_2$ -brine electrolysis operating at terrestrial atmospheric pressure of 101325 Pa was calculated using the expression,

$$E_{Nernst} = |E_{emf}| - \frac{RT}{8F} ln\left( \frac{[y_{CH_4}][y_{O_2}]^2}{[y_{CO_2}]} \right) \tag{23}$$

where the mole fractions of $CH_4$, $O_2$ and $CO_2$ were written as $y_{CH_4}$ , $y_{O_2}$ and $y_{CO_2}$. $y_{H_2O}$ was ~1 as it was in the liquid phase.

The Nernst potentials of $CO_2$-brine electrolysis was calculated after incorporating the corrections for surface pressure on Mars as[36],

$$E_{Nernst} = |E_{emf}| - \frac{RT}{8F} ln\left( \frac{[y_{CH_4}][y_{O_2}]^2}{[y_{CO_2}]} \times \left[\frac{P_{Mars}}{P_{Std}}\right]^2 \right) \tag{24}$$

The $E_{emf}$ for both the brine and $CO_2$ – brine electrolyzers was determined from the thermodynamic parameters as discussed in the thermodynamic model deveopment section. The contributions from concentration (activity) of electroactive species was calculated by varying the extent of the reaction (conversion) for the reactions from 5 % to 95 % as discussed below.

For example, the Nernst potentials of $CO_2$-brine electrolysis (**eqn. (24)**) was expressed in terms of conversion ($\varepsilon$) as,

$$E_{Nernst} = |E_{emf}| - \frac{RT}{8F} ln\left( \frac{\left[\frac{\varepsilon}{3}\right]\left[\frac{2\varepsilon}{3}\right]^2}{\left[\frac{1-\varepsilon}{3}\right]} \times \left[\frac{P_{Mars}}{P_{Std}}\right]^2 \right) \tag{25}$$

Where the mole fractions of $y_{CH_4}$ , $y_{O_2}$ and $y_{CO_2}$ and $y_{H_2O}$, are $\left[\frac{\varepsilon}{3}\right]$, $\left[\frac{2\varepsilon}{3}\right]$and $\left[\frac{1-\varepsilon}{3}\right]$ respectively. The Nernst potential with 50 % conversion ($\varepsilon$ = $0.5$) at 298K and 636 Pa was calculated to be ~ 1.113 V as shown below,

$$E_{Nernst} = 1.07 - \frac{8.314 \times 298}{8 \times 96485} ln\left(\frac{\left[\frac{0.5}{3}\right]\left[\frac{1}{3}\right]^2}{\left[\frac{0.5}{3}\right]} \times \left[\frac{636}{101325}\right]^2\right) = \sim 1.11\, V \qquad (26)$$

A 3-dimensional (3D) plot for the Nernst potentials at various reaction extents (conversion, $\varepsilon$) and temperatures ranging from 230K to 298K were calculated and plotted as surface plots as shown in **Fig. 3** for brine electrolysis (**Fig. 3(a)**) and $CO_2$-brine electrolysis (**Fig. 3(b)**) at 636Pa and 101325Pa. The various configurations of the electrolyzer, including balance of plant components, for operations at combinations of both terrestrial and Martian average surface pressures and temperatures are depicted in **Fig. S1-S3**. Based on these calculations, we conclude that temperature exhibits a greater impact on the Nernst potentials as compared to pressure and temperature effects will be emphasized going forward.

A comparison of Nernst potentials under Martian conditions (236K, 636 Pa) and terrestrial conditions (298K, 101325 Pa) at various ε values for brine electrolysis and $CO_2$-brine electrolysis is shown in **Fig. 4**. $CO_2$ electrolysis to carbon monoxide (CO) and $O_2$, (the MOXIE reaction) was also studied at low temperatures for comparison. The results clearly show the advantage (lower Nernst potential, hence a lower thermodynamic barrier) of operating a $CO_2$-brine electrolyzer to produce $CH_4$ + $O_2$ compared to a brine electrolyzer to produce $H_2$ + $O_2$ or MOXIE to produce CO + $O_2$ (at low temperatures) under both Martian and terrestrial conditions. It was observed that pressure variation between 636 Pa and 101325 Pa had minimal effect on the performance of the electrolyzer modeled at 236 K.

**Polarization model development**

The polarization model is built to capture the realistic operational voltage of the electrolyzer ($E_{cell}$) working under Martian conditions combining thermodynamic (electromotive force ($E_{emf}$) and Nernst potential ($E_{Nernst}$) for electrolysis along with activation overpotential for electrodes, anode ( $\eta_{act}^{anode}$) and cathode ($\eta_{act}^{cathode}$) and transport losses ((Ohmic losses ($E_{\mathrm{IR}}$) due to electronic and ionic resistances) . The polarization model is thus the sum of all these contributions as shown below-

$$E_{cell} = E_{Nernst} + |\eta_{act}^{anode}| + |\eta_{act}^{cathode}| + E_{IR} \quad (27)$$

$E_{Nernst}$ was calculated from $E_{emf}$ and the reaction stoichiometry using the Nernst equation[36] as detailed in the previous section. Activation overpotential ($\eta$) was determined from linear sweep voltammetry (LSV) curves previously reported by Sankarasubramanian and co-workers.[5]

In addition to the electrode catalyst activation overpotentials $\eta_{act}^{anode}$ and $\eta_{act}^{cathode}$, increases in faradaic (electrochemical reaction) current will also require an applied overpotential as described by the Butler-Volmer equation (**eqn 28**),

$$j = j_0 * \left( exp^{\left(\frac{\alpha_a nF\eta}{RT}\right)} - exp^{\left(\frac{\alpha_c nF\eta}{RT}\right)} \right) \quad (28)$$

Where, $j$ is the current density (mA cm$^{-2}$), $j_0$ is the exchange current density (mA cm$^{-2}$) which is a common metric of electrocatalyst activity, $n$ is the number of electrons, $F$ is Faraday's constant (C mol$^{-1}$), $R$ is the universal gas constant (J K$^{-1}$mol$^{-1}$), $\alpha_a$ and $\alpha_c$ are the dimensionless anodic and cathodic charge transfer coefficients respectively, $\eta$ is the applied overpotential ($E - E_{eq}$), and $T$ is temperature (K). Given that the faradaic current is proportional to the exponent of the overpotential, the overpotential contribution from pulling current is quite small. On the other hand, large overpotentials are needed to overcome the cell resistance (voltage and resistance are directly proportional) and hence the ohmic overpotential contributions are significantly larger. Thus, our model will accurately describe systems using most active catalysts with relatively high exchange current densities.

The polarization model was applied to a well-known water electrolysis reaction (**eqn. (3)**) as an initial validation exercise.

The corresponding Nernst equation for water electrolysis reaction is,

$$E_{Nernst} = |E_{emf}| - \frac{RT}{2F} ln\left(\frac{[H_2][O_2]^{\frac{1}{2}}}{[H_2O]}\right) \tag{29}$$

In the case of water electrolysis, the $|E_{emf}|$ is 1.23V at 298K and 101325 Pa[37] based on the difference between the equilibrium potentials for the oxygen evolution reaction (OER) (**eqn. (5)**) and the hydrogen evolution reaction (HER) (**eqn. (6)**). Accounting for 90% conversion, the $E_{Nernst}$ was calculated to be 1.28V at 298K and 101325Pa. Activation overpotentials for the electrode, anode ( $\eta_{act}^{anode}$) and cathode ($\eta_{act}^{cathode}$) at 298K and 101325Pa were obtained from the difference between the thermodynamic Nernst potentials for the anode ($E_{ner}^{anode}$ ) or cathode ($E_{ner}^{cathode}$ ) and the onset potential for the anode ($E_{onset}^{anode}$ ) or cathode ($E_{onset}^{cathode}$ ) as shown in **eqn. (30)**. The onset potentials of the electrodes were measured using linear sweep voltammetry experiments performed at the respective electrodes.

$$\eta_{act}^{anode} = E_{onset}^{anode} - E_{ner}^{anode} \qquad \eta_{act}^{cathode} = E_{onset}^{cathode} - E_{ner}^{cathode} \tag{30}$$

At standard temperature and pressure (STP) of 298K, 101325 Pa and pH =0, the activation overpotentials for the OER at anode ($\eta_{act}^{anode}$) and the HER at cathode ($\eta_{act}^{cathode}$) were calculated from their half-cell Nernst potentials as,

$$\eta_{act}^{anode} = E_{onset}^{anode} - 1.28 \qquad\qquad \eta_{act}^{cathode} = E_{onset}^{cathode} - 0 \tag{31}$$

Having obtained the thermodynamic contributions to overall electrolyzer polarization, we turned to the Ohmic losses in the electrolyzer. The most resistive component (and hence the major contributor to Ohmic loses) is the resistance of the membrane, which was calculated using Ohm's law-

$$E_{IR} = jASR_{mem} \tag{32}$$

where, j is the current density (mA $cm^{-2}$) and $ASR_{mem}$ is the area specific resistance (ASR) of the membrane (Ω $cm^2$).

The polarization model applied to both the Martian brine electrolyzer and the $CO_2$-$H_2O$ electrolyzer and for both the cell and stack configurations has accounted for inefficiencies at several levels to provide realistic predictions of performance. First, a reactant conversion of 90% is assumed in the Nernst equation, reducing the operating potential of the electrolyzer. Secondly, the practically achievable open circuit potential (OCP) (which should ideally equal the Nernst potential) is assumed to be a mixed potential with a value equal to 70% of the calculated Nernst potential to account for potential contributions from unwanted side-reactions. These inefficiencies increase the electrolyzer power requirements for a desired production rate (which is proportional to current). Furthermore, catalyst faradaic efficiency (values assumed are a function of the catalyst and are listed for every cell configuration modeled) at each electrode reduces the desired product production rate for a given operating current. Finally, accounting for resistive losses at the electrolyzer cell and stack level, we have modeled the production rates at various electrolyzer efficiencies. Thus, the model conservatively builds in four different inefficiencies and this approach results in a very close model fit with our previously published Martian $H_2O$ electrolysis data as detailed in the next section.

## Model validation

The polarization model expression was validated against experimental data from our earlier studies on Martian brine $H_2$-$O_2$ electrolyzer before applying the model to the $CO_2$-brine electrolyzer.

Some key assumptions underlying the operation of the brine electrolyzer are -

- The brine solution is made up of 2.8M $Mg(ClO_4)_2$ dissolved in water and is assumed to be in liquid phase at 237K and 101325Pa
- The brine solution is acidic with pH = 3 as measured at standard atmospheric conditions
- Two configurations of the brine electrolyzer were considered –
    - **$Pb_2Ru_2O_7$ Configuration**: OER is catalyzed at the anode by lead ruthenate pyrochlore ($Pb_2Ru_2O_7$) and HER is catalyzed at the cathode by platinum on carbon (Pt-C)
    - **$RuO_2$ Configuration**: OER is catalyzed at the anode by ruthenium oxide ($RuO_2$) and HER is catalyzed at the cathode by platinum on carbon (Pt-C)
- The onset potentials for the $Pb_2Ru_2O_7$ / $RuO_2$ anode ($E_{onset}^{anode}$ ) and Pt-C cathode ($E_{onset}^{cathode}$ ) were calculated from the experimental values from our previous work[5].
- A commercial Fumasep® FAA-3-50 anion exchange membrane (AEM) separator was used in our previous work and its reported ASR was used in our electrolyzer validation calculations.

To develop this polarization model, first $E_{\mathrm{Nernst}}$ for brine electrolysis at various temperatures (~ 230-500K) and pressures (636Pa and 101325Pa) was calculated using the thermodynamic data and the Nernst equation. Second, the activation overpotentials at the anode and cathode were calculated at the specific operating conditions of the brine electrolyzer. Finally, the Ohmic losses were calculated from the membrane resistance ($R_{mem}$) at varying *i* values.

The $E_{Nernst}$ for a brine electrolyzer operating at 237K and 101325Pa with a conversion efficiency of 90% was calculated to be

$$E_{Nernst} = 1.34V \tag{33}$$

The activation losses differed based on the electrolyzer configuration (due to variations in electrocatalytic characteristics of the catalysts). The calculations are detailed below -

- **$Pb_2Ru_2O_7$ anode and Pt-C cathode Configuration:** The activation overpotential for OER on $Pb_2Ru_2O_7$ anode and HER on Pt-C cathode are calculated from half-cell Nernst potentials at pH = 0,

$$H_2O = \frac{1}{2}O_2 + 2H^+ + 2e^- \qquad (E_{ner}^{anode,237k,101325Pa} = -1.3V) \tag{34}$$

$$2H^+ + 2e^- = H_2 \qquad (E_{ner}^{cathode,237k,101325Pa} = 0V) \tag{35}$$

Correcting it to electrolyzer operating conditions of 237 K, 101325 Pa with pH = 3 as,

$$H_2O = \frac{1}{2}O_2 + 2H^+ + 2e^- \qquad (E_{ner}^{anode,237k,101325Pa} = -1.16V) \tag{36}$$

$$2H^+ + 2e^- = H_2 \qquad (E_{ner}^{cathode,237k,101325Pa} = -0.14V) \tag{37}$$

$$|\eta_{act}^{Pb_2Ru_2O_7}| = E_{onset}^{Pb_2Ru_2O_7} - 1.16 \qquad |\eta_{act}^{Pt-C}| = E_{onset}^{Pt-C} - (-0.14) \tag{38}$$

$E_{onset}^{Pb_2Ru_2O_7}$ and $E_{onset}^{Pt-C}$ for the anode and cathode were calculated to be 1.35V and -0.55V from the experiments performed on brine at 237K and 101325Pa. The calculated activation overpotential for the electrodes were,

$$|\eta_{act}^{Pb_2Ru_2O_7}| = 0.19V \; and \; |\eta_{onset}^{Pt-C}| = 0.41V \tag{39}$$

The total activation overpotential was calculated to be

$$\eta_{act} = |\eta_{act}^{Pb_2Ru_2O_7}| + |\eta_{act}^{Pt-C}| = 0.6V \tag{40}$$

The $R_{mem}$ for the Fumasep® FAA-3-50 membranes used were reported to vary between 0.2 - 0.7 Ω $cm^2$ [38]. We have conservatively used the maximum value of 0.7 Ω $cm^2$ for calculating $E_{\Omega}$ in our model. The predicted operating cell potentials of the brine electrolyzer (solid red line) at 237K and 101325Pa with $Pb_2Ru_2O_7$ anode and Pt-C cathode is shown in **Fig. 5**. The experimental values (brown squares) are compared with the calculated operating cell potentials (solid red line) as shown in **Fig. 5**. The experimental values lie close to the 70% efficiency (dotted red line) of the calculated values of the model. Given that the state-of-the-art electrolyzers operate at 70–80% electricity-to-hydrogen efficiency to produce high-purity (>99.9%) $H_2$ [39], the close match with predicted performance at 70% efficiency serves as a powerful validator of our model. The model was also applied to configuration 2 as detailed below.

- **$RuO_2$ anode and Pt-C cathode Configuration:** The activation overpotential for the OER at $RuO_2$ anode and the HER at Pt-C cathode are calculated from half-cell Nernst potentials which have the same values presented in the previous section. The activation overpotentials were calculated as -

$$|\eta_{act}^{RuO_2}| = E_{onset}^{RuO_2} - 1.16 \qquad |\eta_{act}^{Pt-C}| = E_{onset}^{Pt-C} - (-0.14) \tag{41}$$

$E_{onset}^{RuO_2}$ and $E_{onset}^{Pt-C}$ for the anode and cathode were calculated to be 1.55V and -0.55V from the experiments performed with brine at 237K and 101325Pa. The calculated activation overpotential for the electrodes were,

$$|\eta_{act}^{RuO_2}| = 0.39V \qquad |\eta_{act}^{Pt-C}| = 0.41V \tag{42}$$

The total activation overpotential was calculated to be

$$\eta_{act} = |\eta_{act}^{RuO_2}| + |\eta_{act}^{Pt-C}| = 0.80V \tag{43}$$

As the separator membrane did not change between the configurations, the Ohmic losses remained the same as in configuration 1. A model representing the operating cell potentials of the brine electrolyzer (red line) at 237K and 101325Pa with $RuO_2$ anode and Pt-C cathode is shown in **Fig. 6**. The polarization performance at 70% efficiency is depicted using a dotted red line and is expected to closely match the performance of an actual electrolyzer with this configuration. $RuO_2$ was used in the polarization calculations for electrolyzer modeling due to its degradation resistance and compatibility under acidic conditions as discussed in our previous study[40] and due to availability of overpotential data at operating conditions of the electrolyzer i.e. 255 K and in a $Mg(ClO_4)_2$ electrolyte. The $CO_2$-brine electrolyzer was also modeled using a $RuO_2$ anode as detailed in the next section.

**$CO_2$-brine electrolysis to produce $CH_4$ and $O_2$ on Mars**

The key assumptions underlying the operation of the $CO_2$-brine electrolyzer are -

- The brine solution is made up of 2.8 M $Mg(ClO_4)_2$ dissolved in water and is assumed to be in liquid phase at 255K and 101325Pa (see detailed discussions on the thermodynamic validity of this assumption in our previous work[5])
- The brine solution is acidic with pH = 3 as measured at standard atmospheric conditions.

- OER is carried out at the anode by ruthenium oxide ($RuO_2$) and carbon-dioxide reduction reaction ($CO_2RR$) is carried out at the cathode by copper (Cu(111)) and the activation overpotentials are calculated from the onset potentials of experiments with brine at 255K and 101325Pa. For $CO_2$-brine electrolysis, 255 K and 101325 Pa is considered for model instead of 236 K and 101325 Pa because, the Cu (111) cathode ($E_{onset}^{cathode}$ ) values are reported at 255 K in the literature[26].

We calculated $E_{Nernst}$ for the $CO_2$-brine electrolysis at various temperatures (~ 230 - 500K) and pressures (636Pa and 101325Pa) using the thermodynamic data and the Nernst equation. The $E_{Nernst}$ for the $CO_2$-brine electrolyzer operating at 255K and 101325Pa with a conversion efficiency of 90% is,

$$E_{\text{Nernst}} = 1.08\text{V} \tag{44}$$

The half-cell Nernst potentials at 237 K, 101325 Pa and pH = 0 are -

$$4H_2O = 2O_2 + 8H^+ + 8e^- \qquad (E_{ner}^{anode,237k,101325Pa} = -1.30V) \tag{45}$$

$$CO_2 + 8H^+ + 8e^- = CH_4 + 2H_2O \qquad (E_{ner}^{cathode,237k,101325Pa} = 0.22V) \tag{46}$$

Correcting to pH = 3, we obtained --

$$4H_2O = 2O_2 + 8H^+ + 8e^- \qquad (E_{ner}^{anode,237k,101325Pa} = -1.14V) \tag{47}$$

$$CO_2 + 8H^+ + 8e^- = CH_4 + 2H_2O \qquad (E_{ner}^{cathode,237k,101325Pa} = 0.058V) \tag{48}$$

$$|\eta_{act}^{RuO_2}| = E_{onset}^{RuO_2} - 1.14 \qquad \left|\eta_{act}^{Cu(111)}\right| = E_{onset}^{Cu(111)} - (0.058) \tag{49}$$

$E_{onset}^{RuO_2}$ and $E_{onset}^{Cu(111)}$ for the anode and cathode were calculated to be 1.49V and -0.25V from the experiments performed with brine at 255K and 101325Pa. The calculated activation overpotential for the electrodes are,

$$|\eta_{act}^{RuO_2}| = 0.34V \qquad |\eta_{act}^{Cu(111)}| = 0.32V \tag{50}$$

The total activation overpotential is thus-

$$\eta_{act} = |\eta_{act}^{RuO_2}| + |\eta_{act}^{Cu(111)}| = 0.66V \tag{51}$$

The Ohmic loss contributions were parameterized in our model of a $CO_2$-brine electrolyzer by assuming the conductivity of the separator membrane for the $CO_2$-brine electrolyzer to be equal to the conductivity of Nafion 117 at 303K. Thus, $ASR_{mem}$ is estimated to be $\sim$ 1.22 Ω $cm^2$ [41] . A higher conductivity separator membrane is thus a critical requirement for further improving the performance of these electrolyzer systems. The model predictions of the operating cell potentials of the $CO_2$-brine electrolyzer (red line) at 255K and 101325Pa with $RuO_2$ anode and Cu(111) cathode is shown in **Fig. 7**. Nafion 117 was selected for this polarization model due to the availability of data at lower temperatures and compatibility with the acidic conditions present in the electrolyzer. Further development of operational electrolyzers will utilize thinner and less resistive membranes to further increase performance.

**Performance of a $CO_2$-brine electrolyzer stack**

As shown in our model validation for the brine electrolyzer (**Fig. 6**), assuming a practical open circuit potential that is 70% of the calculated Nernst potential closely matches experimental data. Thus, this validated assumption was taken as the operational single-cell potential of the $CO_2$-brine electrolyzer and appropriately scaled-up (based on the assumed series- and parallel configuration of the cells in the stack) for our stack calculations. The electrolyzer stack was assumed to contain 10 cells of 100 $cm^2$ electrode surface area (each cell) and its performance is shown in **Fig. 8**. The operational cell potential of the stack along with the power requirements are shown in the primary Y-axis and the volume of $CH_4$ and $O_2$ generated by the stack is shown as secondary Y-axis with the current density as the X-axis. The $CH_4$ generation rate (for a given electrolyzer operating power) was calculated assuming a cathode faradaic efficiency of 36% (matching the reported efficiency for the Cu catalyst[26]) while the anode production rates of $O_2$ were calculated assuming a anodic faradaic efficiency of 70% (matching the reported efficiency for the $RuO_2$ catalyst).

Our electrolyzer system operating with $E_{stack}$ of 20 V (2.0V applied at each cell) has a theoretical power normalized production of ~ 0.45 g $W^{-1}$ $day^{-1}$ of $CH_4$ and 3.55 g $W^{-1}day^{-1}$ of $O_2$ (**Fig. 9**), assuming 50% system efficiency (same as MOXIE[21]). This translates to >1000L of both $CH_4$ and $O_2$ per day using this conservatively sized system[42,43]. Thus, with the appropriate scaling, such an electrolyzer can produce sufficient fuel and oxidant for both payload return and life support for future missions to Mars. Notably, our $CO_2$-brine electrolyzer produces both propellant components for methalox engines anticipated to power future American and International missions to Mars. The critical next steps would be the down selection and experimental screening of efficient and selective catalysts for $CO_2$RR under Martian conditions.

**Catalyst requirements for a Martian $CO_2$-brine electrolyzer**

The model predicts the performance of a $CO_2$-brine electrolyzer employing well-known Cu catalyst for the cathodic $CO_2$RR with varying faradaic efficiencies. We anticipate the greatest improvements in performance to result from the use of alternate, high activity, high selectivity catalysts. Currently, the $RuO_2$ anode's predicted operational oxygen generation efficiency of 70% outperforms elemental copper cathode's operational methane generation efficiency of 36%[26], which would reduce the production rate of oxygen to match. To address this concern, a list of possible catalyst candidates is presented in **Table 3**. Replacing Cu with Cu-Bi or Cu-Zn alloys or single atom Zn is expected to significantly increase (potentially double) the production rate of $CH_4$ in our system given their much higher faradaic efficiency.  Furthermore, the ultra-cold temperatures and liquid brines present on the Martian surface have been shown to increase the methane selectivity of copper $CO_2$RR electrocatalysts[26]. We anticipate this solvation effect to translate to these catalyst candidates as well and further increase their $CH_4$ selectivity and ultimately our electrolyzer.

## Conclusion

A $CO_2$-brine electrolyzer operating at Martian ambient conditions has been found to be a viable and energy efficient solution for fuel and oxygen production on Mars through ISRU. Following successful validation of the model against experimental data, our model predicts our >1000L of $CH_4$ and $O_2$ production using a modestly sized 10-cell stack. The further development of this system incorporating more efficient $CO_2$RR catalysts and higher conductivity separators is expected to significantly improve performance and present a viable solution for fuel and oxygen production for sustained human exploration on Mars.

## Acknowledgements:

The authors gratefully acknowledge funding through a start-up grant from the University of Texas at San Antonio.

## Author contributions:

Dr. Mohamed Shahid: Data curation, formal analysis, investigation, methodology, resources, software, validation, visualization, writing – original draft, writing – review & editing.

Mr. Bradley Chambers: Investigation, resources, data curation, validation, visualization, writing – original draft, writing – review & editing.

Dr. Shrihari Sankarasubramanian: Conceptualization, funding acquisition, writing – review & editing, supervision, validation, project administration, resources.

**Conflicts of interest**: The authors declare no competing financial interest.

## Tables

**Table 1.** Comparison of conditions on Mars and on Earth[44].

| | **Earth** | **Mars** |
|---|---|---|
| **Duration of revolution around the sun (days)** | 365.24 | 686.97 |
| **Duration of rotation about its axis (h)** | 24 | 24.66 |
| **Average diurnal surface temperature range (K)** | 283 to 293 | 184 to 242 |
| **Average surface pressure (Pa)** | 101400 | 636 (seasonally variable from 400 – 870) |
| **Surface gravity ($m.s^{-2}$)** | 9.79 | 3.71 |
| **Escape Velocity km/s** | 11.19 | 5.03 |
| **Atmospheric composition** | *Major (vol.%):* 78.08% $N_2$, 20.95% $O_2$, >1% $H_2O$ (highly variable)<br>*Minor:* 9340 ppm Ar, 410 ppm $CO_2$, 18.18 ppm Ne, 5.24 ppm He, 1.7 ppm $CH_4$, 1.14 ppm Kr, 0.55 ppm $H_2$ | *Major (vol.%):* 95.1% $CO_2$, 2.59% $N_2$, 1.94% Ar, 0.16% $O_2$, 0.06% CO<br>*Minor:* 210 ppm $H_2O$, 100 ppm NO, 2.5 ppm Ne, 0.85 ppm H-D-O, 0.3 ppm Kr, 0.08 ppm Xe |

**Table 2.** Representative methalox engines in use or development around the World[45].

| Engine Name | HD | MIRA | ACE-42R | BE-4 | Raptor | LE-8 | AEON-1 | TQ-12 |
|---|---|---|---|---|---|---|---|---|
| Country | USA | Italy/EU | France | USA | USA | Japan | USA | China |
| Company | NASA | Avio | Airbus | Blue Origin | SpaceX | JAXA | Relativity Space | Landspace |
| Thrust kN | 24 | 98 | 420 | 2400 | 2200 | 107 | 100 | 670 |
| $O_2/CH_4$ Ratio | 3.4-3.8 | 3.4 | 3.4-3.8 | 3.4-3.8 | 3.6 | 3.4-3.8 | 3.4-3.8 | 3.4-3.8 |

**Table 3:** Recently reported $CO_2$RR catalysts

| Catalyst | Products | Faradaic Efficiency | Temperature | Ref. |
|---|---|---|---|---|
| CoPc on Zn-N-C | $CH_4$/CO | 18.3% $FE_{CH_4}$ | 298K | 46 |
| Cu with 15 µm of Nafion | $CH_4$/CO | 88% $FE_{CH_4}$ | RT | 47 |
| Pt | $CH_4$/CO | 1.19% $FE_{CH_4}$ | 303K | 48 |
| CuAl Ga doped | $CH_4$/CO/ $C_2H_4$ | 54% $FE_{CH_4}$ | RT | 49 |
| Single Atom Zn-MNC | $CH_4$/CO | 85% $FE_{CH_4}$ | RT | 50 |
| $Cu_{.7}Bi_{.3}$ bimetallic | CH4/CO/ HCOOH | 70.6% $FE_{CH_4}$ | RT | 51 |
| Cu Oh-NC 75nm | $CH_4$/$C_2H_4$/ $HCOO^-$ | 55% $FE_{CH_4}$ | RT | 52 |
| Copper (II) phthalocyanine | CH4 C2H4 HCOOH | 66% $FE_{CH_4}$ | RT | 52 |
| Polycrystalline Cu | $CH_4$/CO/ $H_2$ | 40.4% $FE_{CH_4}$ | RT | 53 |
| CuS Nanosheets | $CH_4$/CO/ $H_2$ | 73.5% $FE_{CH_4}$ | 298K | 54 |
| Pd decorated Cu | $CH_4$/$H_2$ | 50% $FE_{CH_4}$ | RT | 55 |
| Cu doped CuO2 | $CH_4$ | 58% $FE_{CH_4}$ | RT | [56] |
| $Cu_{0.7}Zn_{0.3}$ Catalyst | $CH_4$/CO | 70% $FE_{CH_4}$ | RT | [57] |

**Figures:**

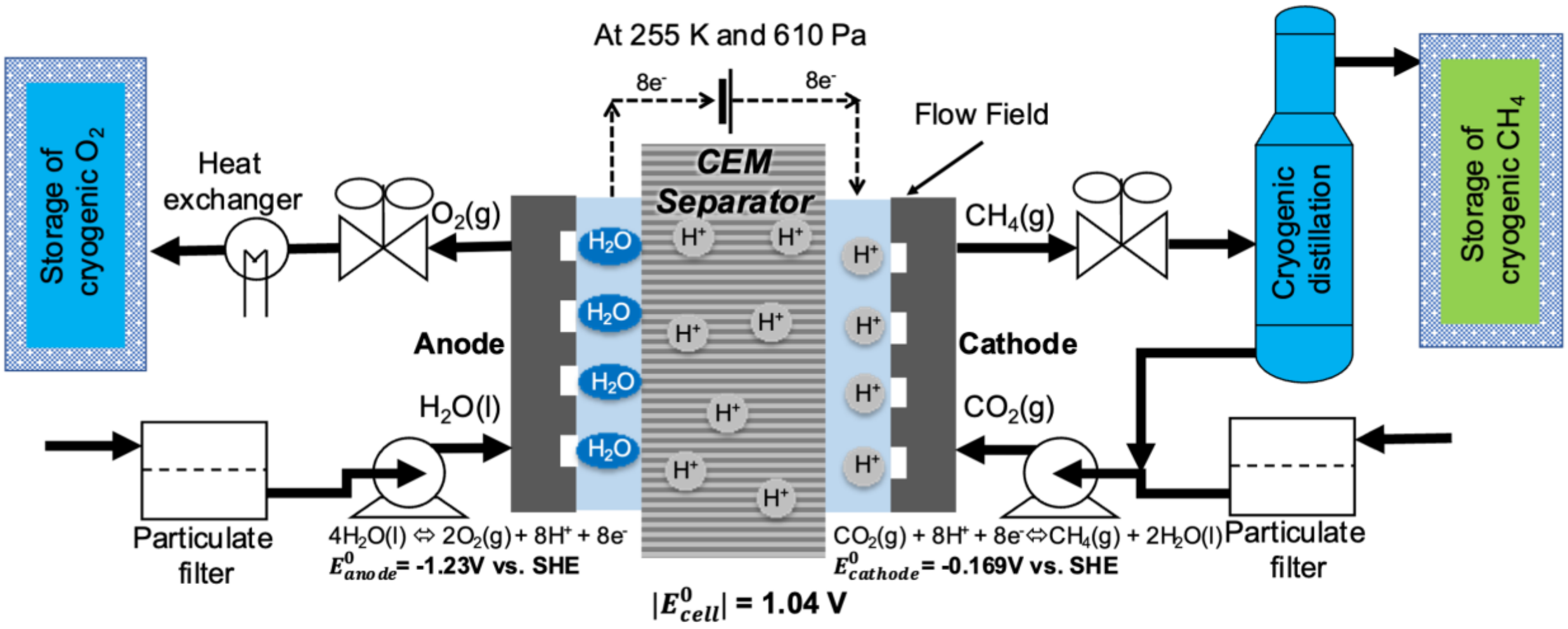

**Figure 1**: Schematic representation of a $CO_2$ -brine electrolyzer for operation on Mars.

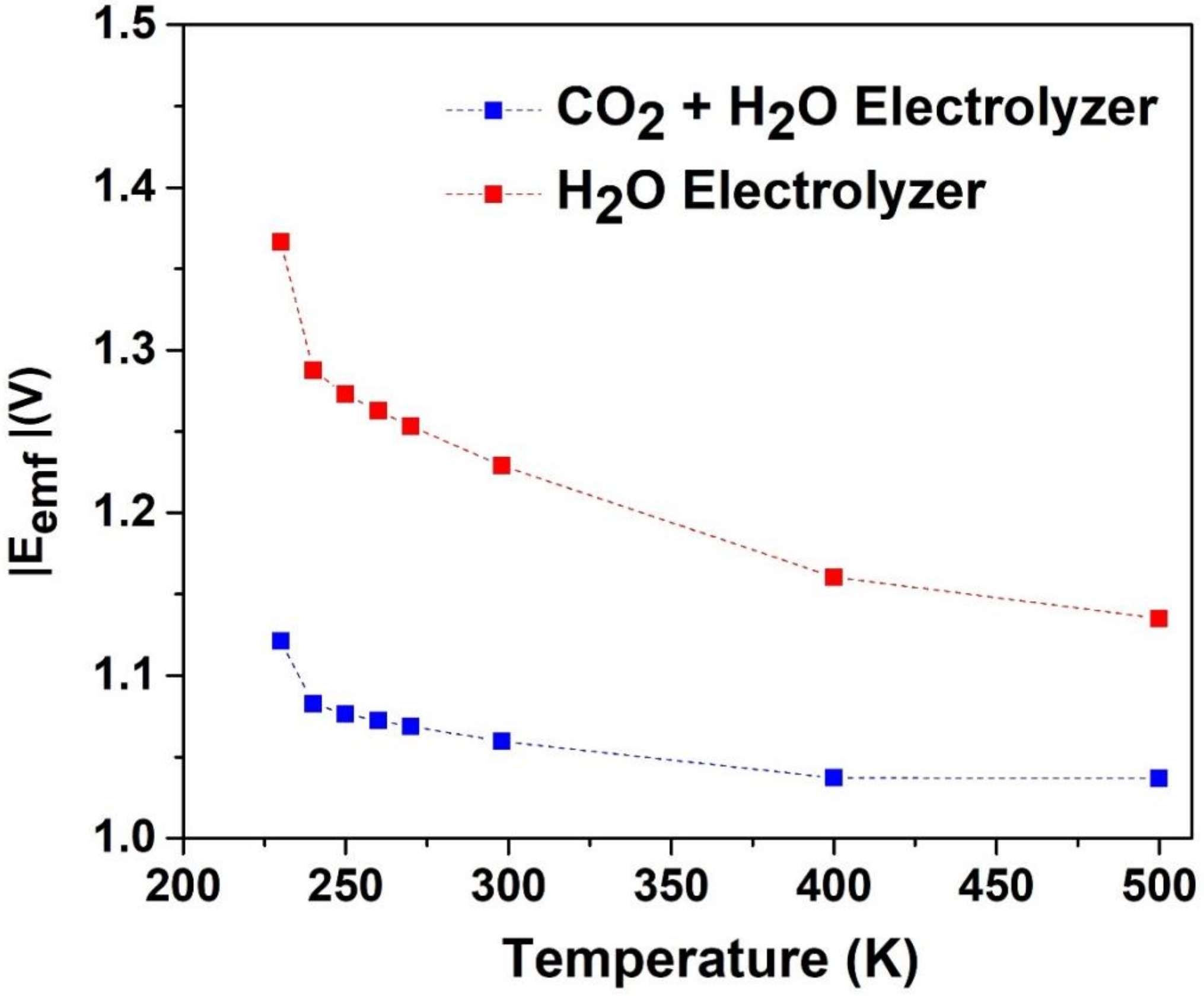

**Figure 2**: The calculated thermodynamic potential, $E_{emf}$, required for brine electrolysis and $CO_2$-brine electrolysis between 230K and 500K.

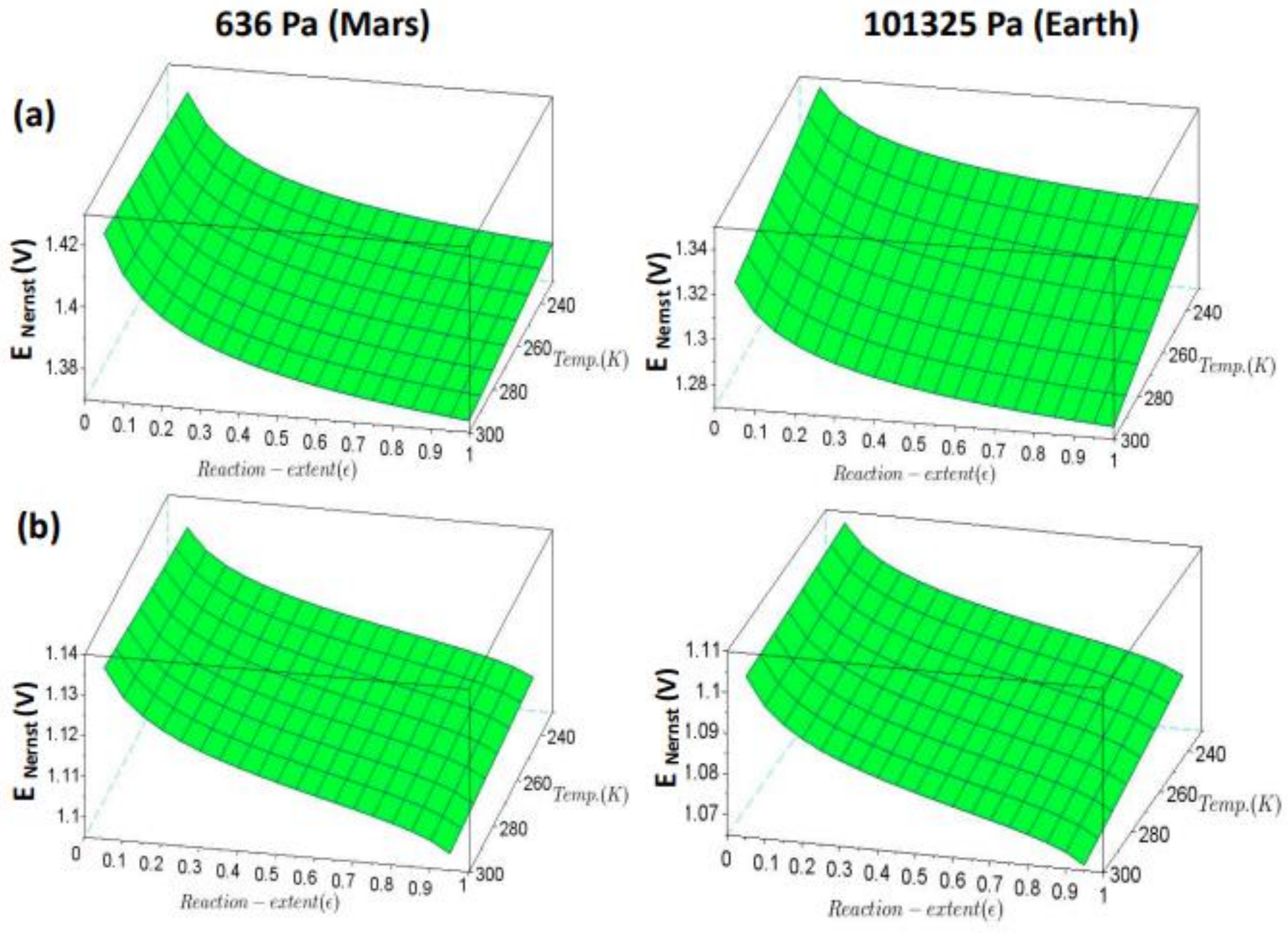

**Figure 3**: Nernst potentials at 636Pa and 101325Pa for (a) brine electrolysis and (b) $CO_2$-brine electrolysis.

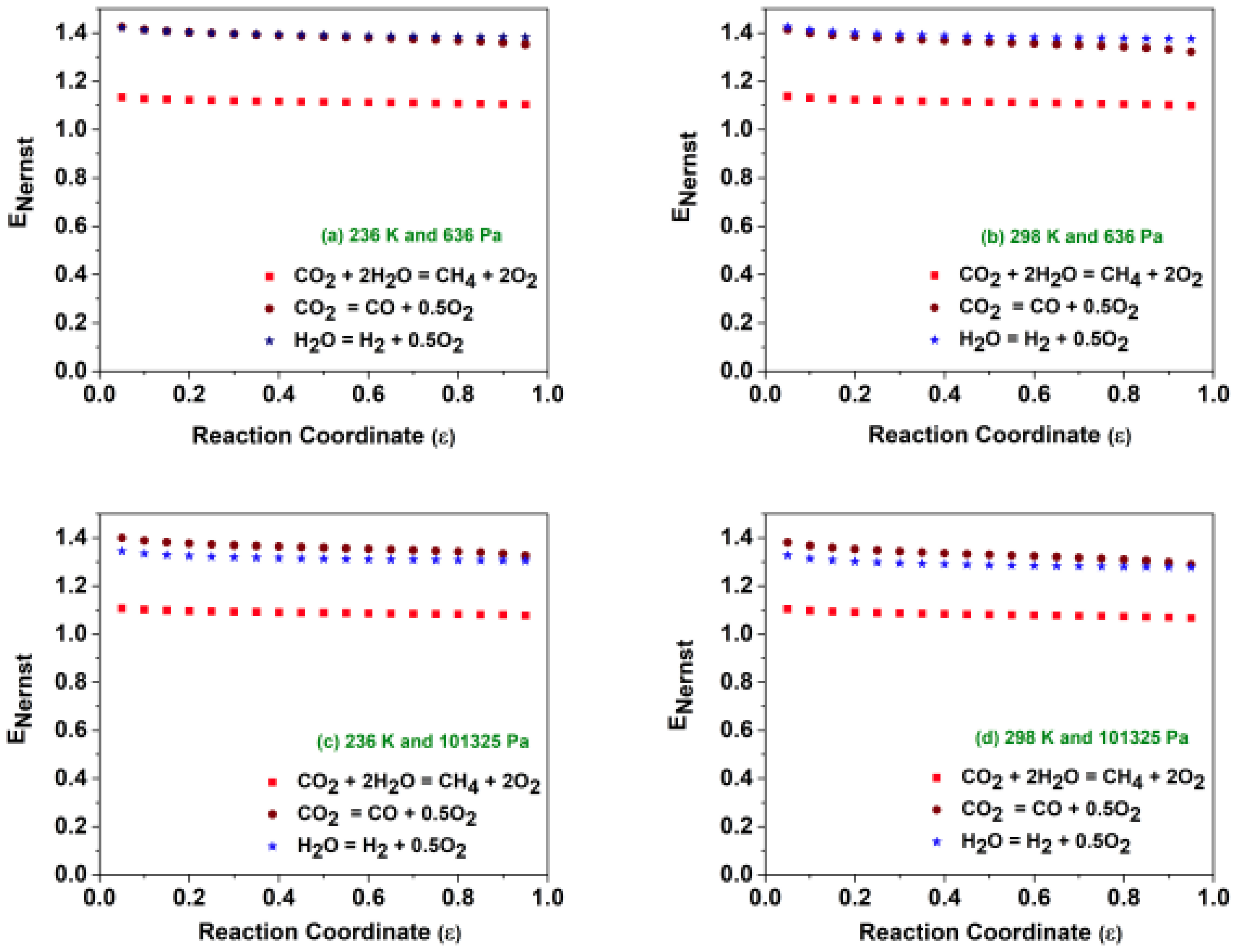

**Figure 4**: Nernst potential for carrying out electrolysis at (a) 236K - 636Pa, (b) 298K - 636Pa, (c) 236K-101325Pa and (d) 298K-636Pa.

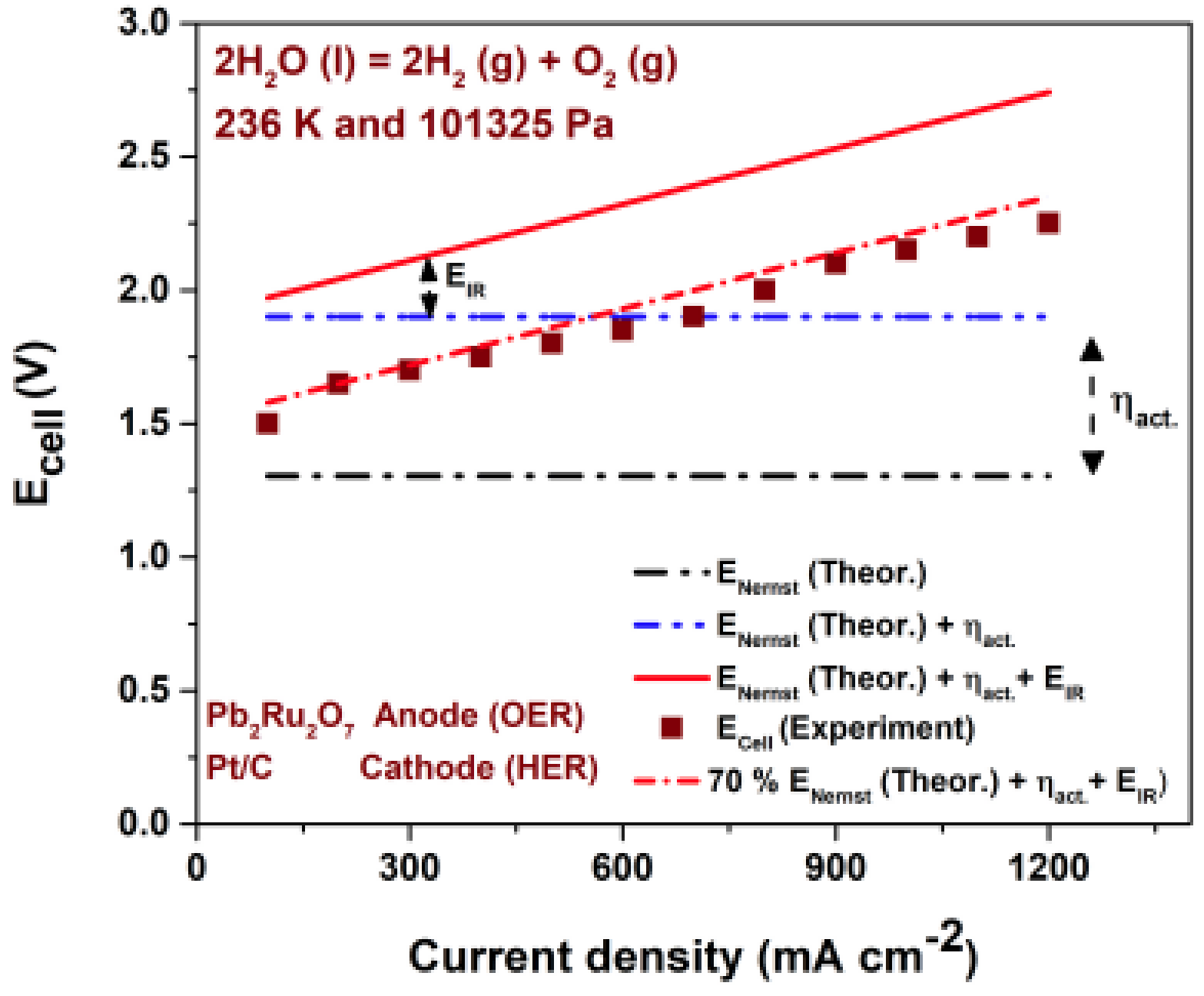

**Figure 5**: Performance of the brine Electrolyzer operating with Pt-C cathode and $Pb_2Ru_2O_7$ anode at 237K and 101325Pa.

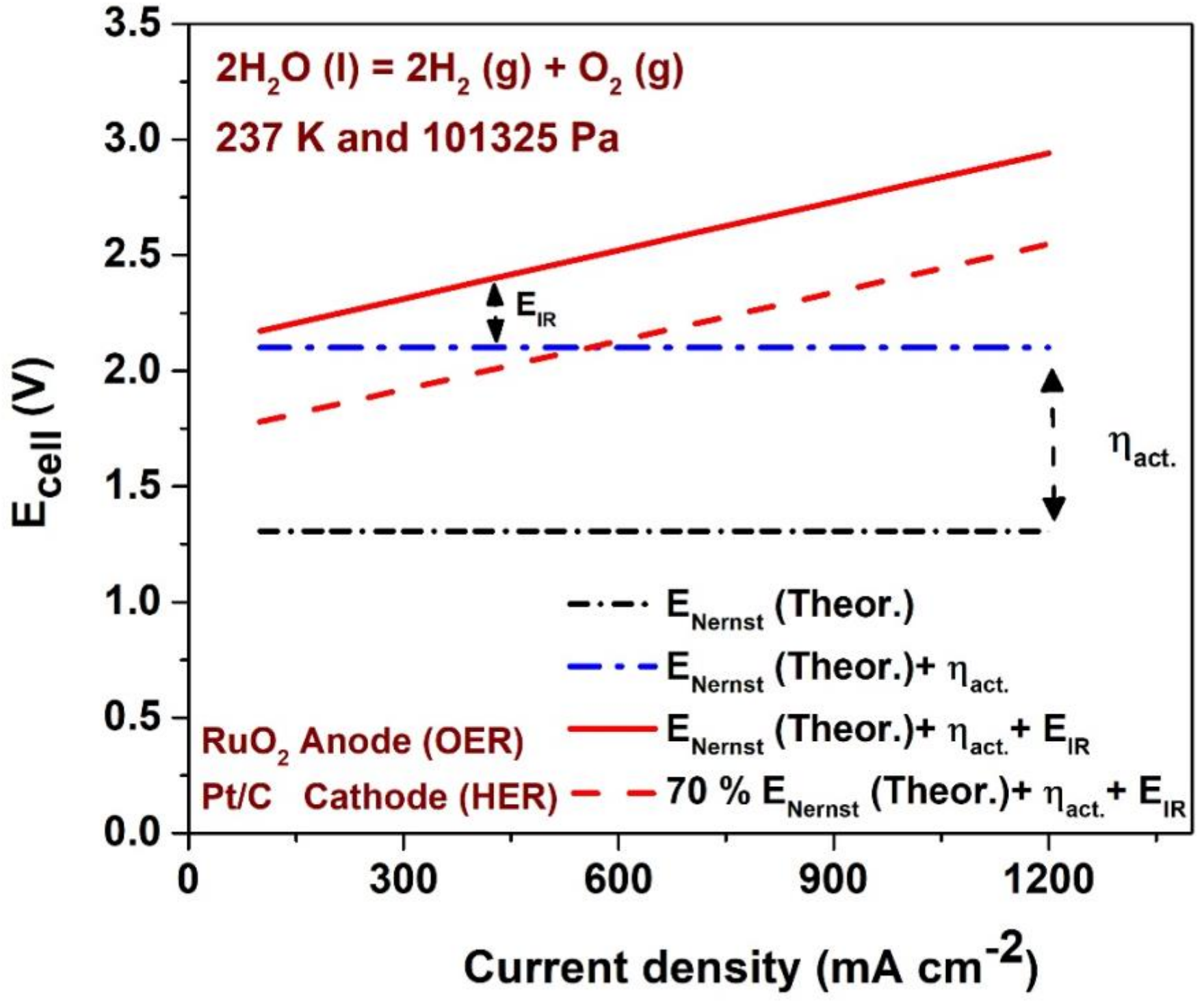

**Figure 6**: Performance of the brine Electrolyzer operating with Pt-C cathode and $RuO_2$ anode at 237K and 101325Pa.

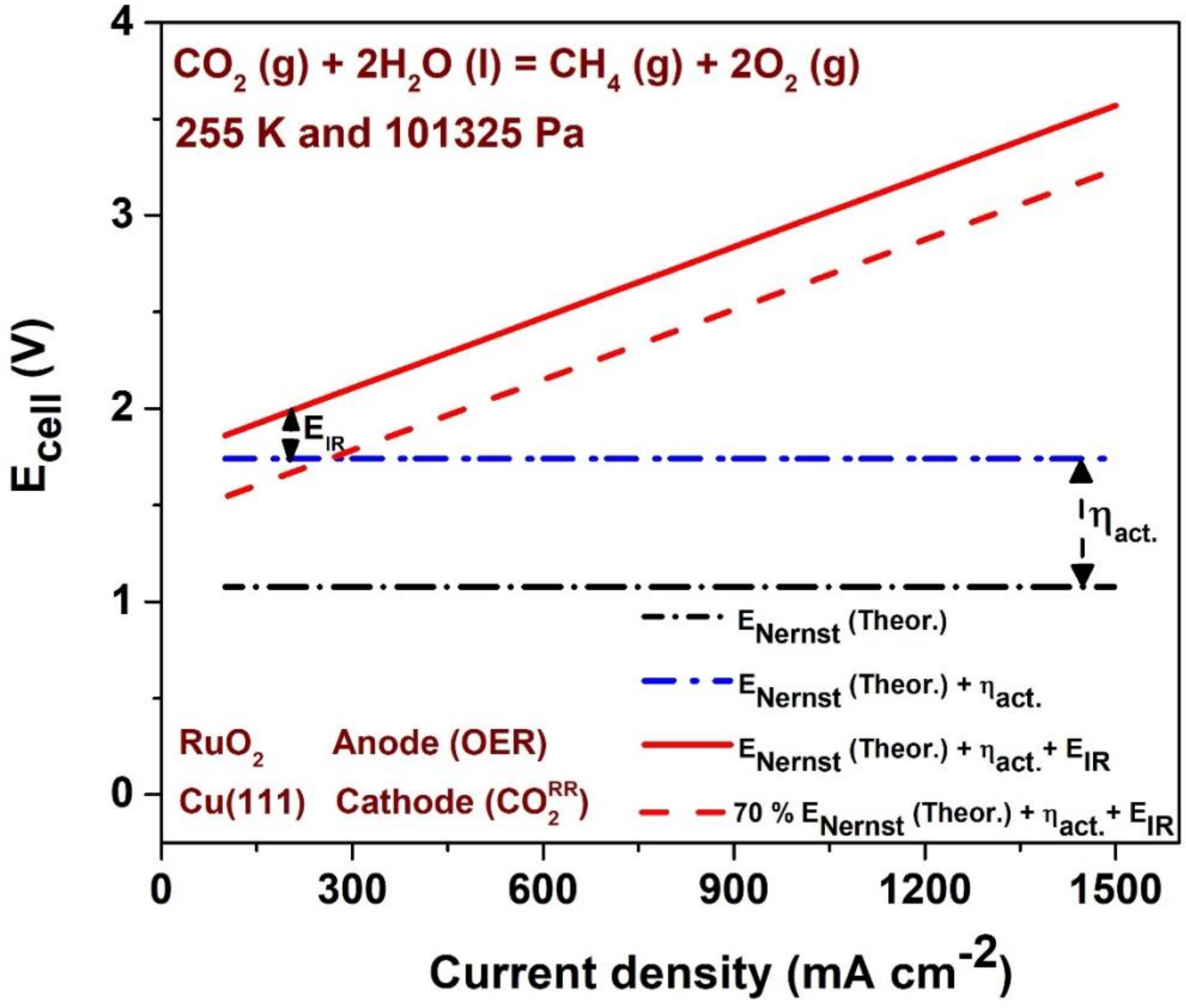

**Figure 7**: Performance of the $CO_2$-brine electrolyzer operating with Cu(111) cathode and $RuO_2$ anode at 255K and 101325Pa.

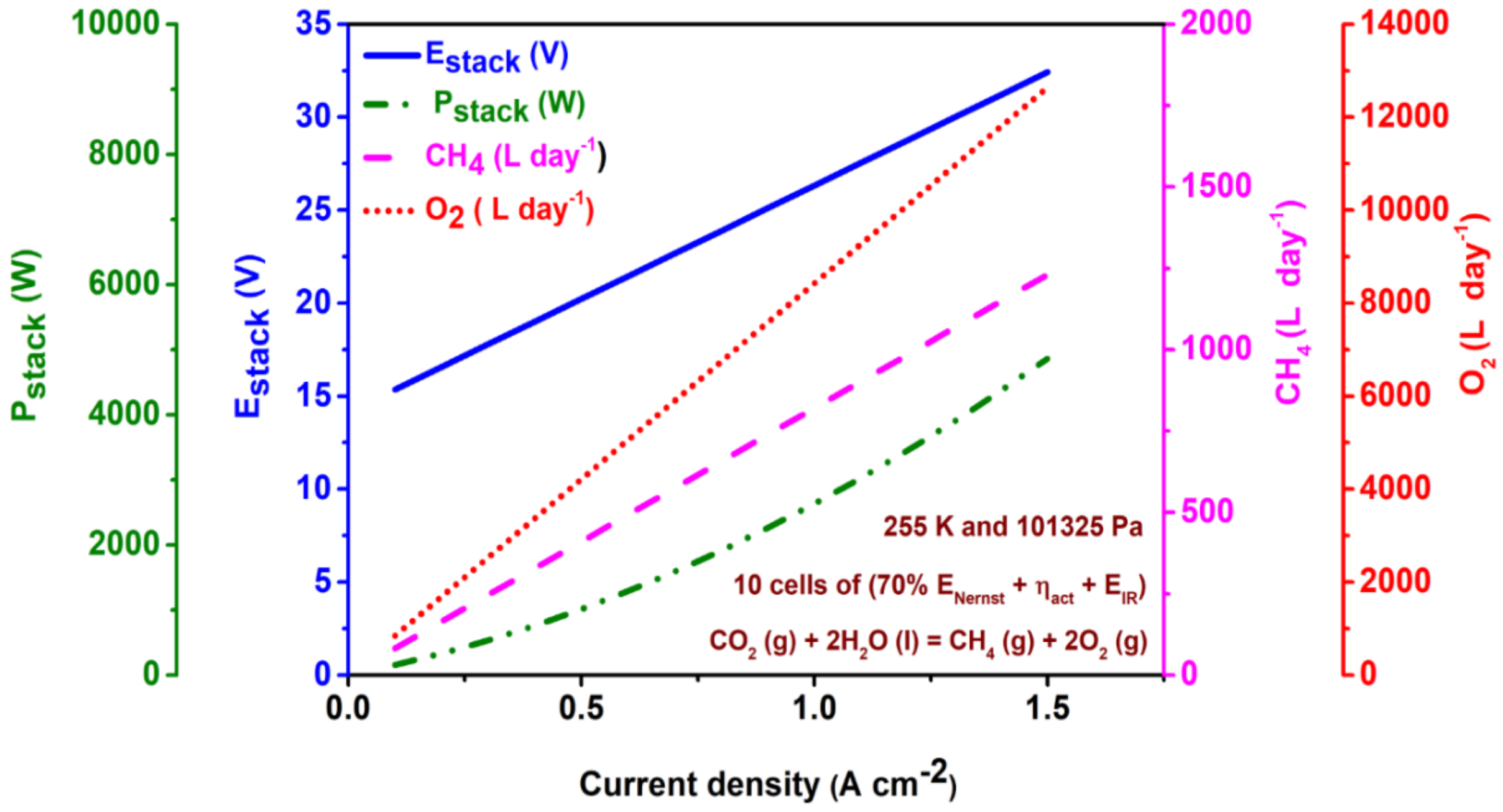

**Figure 8**: Volume of $CH_4$ and $O_2$ produced and power requirements for a stack of $CO_2$-brine electrolyzer operating with 10 cells of 100 cm$^2$ each at 255K and 101325Pa.

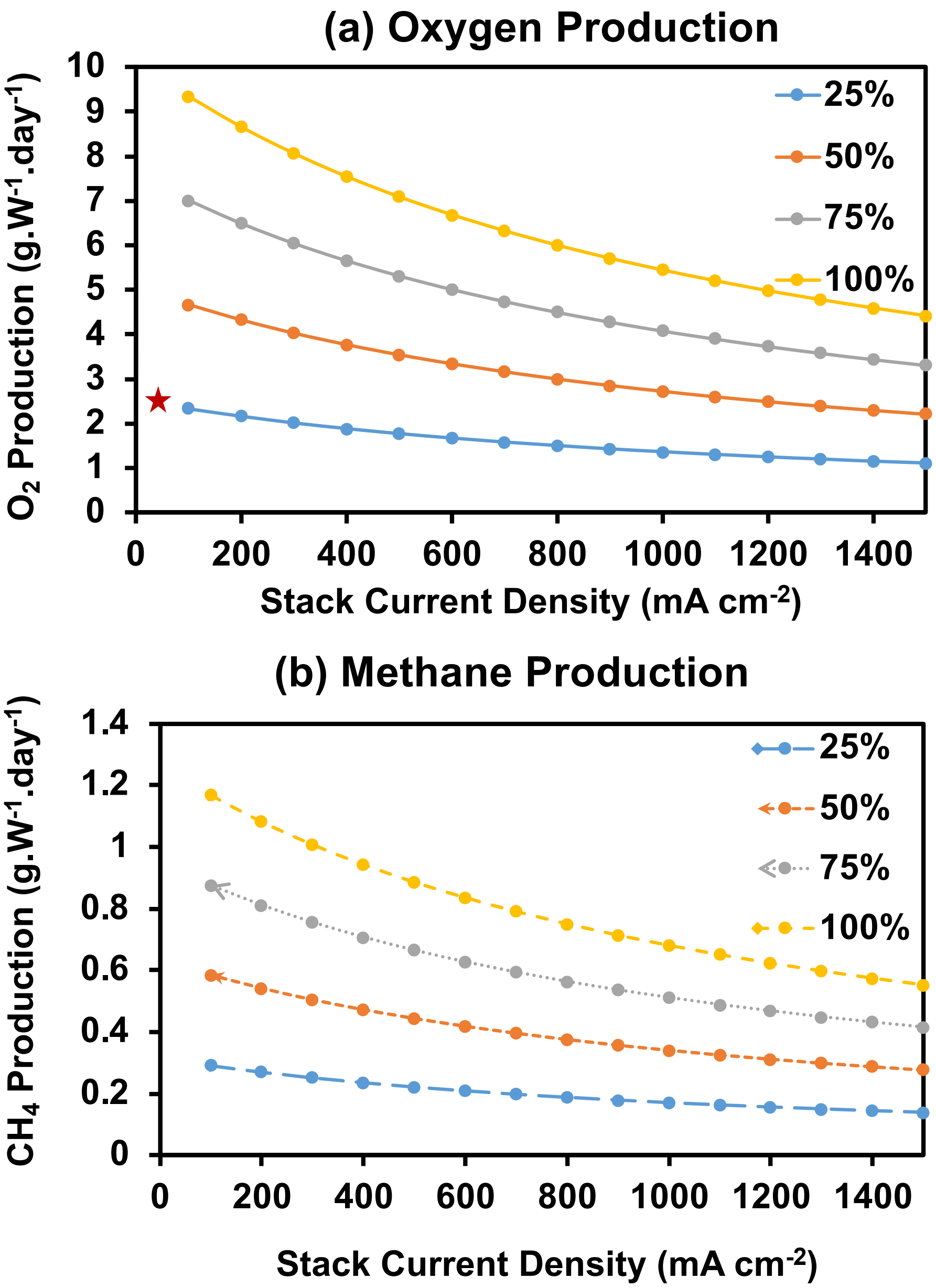

**Figure 9**: Predicted power normalized production rates for (a) oxygen and (b) methane with varying overall system efficiencies (listed in the figure legend) in an $CO_2$-$H_2O$ electrolyzer operating with a stack of 10 cells of 100 cm$^2$ each at 255K and 101325Pa. The best-case production rate for MOXIE is also depicted (red star) for comparison.

**SUPPLEMENTARY INFORMATION**

**For brine-electrolysis**

**$H_2O = H_2 + 1/2O_2$**

The dH, dS and dG values in table S1 are calculated from the[1]

$$C_p = 0.44 * \left(\frac{T - 222}{222}\right)^{-2.5} + 74.3 \qquad (\text{for} - H_2O) \qquad (1)$$

**Table S1: Estimation of $H_2O$ (l) thermodynamic properties for brine electrolysis**

| Temp | dH (kJ mol-1) | dS (kJ mol-1) | dG (kJ mol-1) |
|---|---|---|---|
| 225 | 1804.31 | 6.019 | 450.09 |
| 230 | 412.25 | 0.645 | 263.72 |
| 235 | 323.92 | 0.307 | 251.73 |
| 240 | 303.77 | 0.230 | 248.47 |
| 250 | 293.13 | 0.190 | 245.61 |
| 260 | 290.03 | 0.178 | 243.64 |
| 270 | 288.47 | 0.173 | 241.85 |
| 298 | 285.53 | 0.163 | 237.14 |
| 400 | 242.29 | 0.047 | 223.95 |
| 500 | 243.83 | 0.049 | 219.05 |

**Table S2: Estimation of Thermoneutral voltage and electromotive force from the thermodynamic properties for brine electrolysis.**

| Temp (C) | -48 | -43 | -33 | -23 | -13 | -3 | 25 | 127 | 227 |
|---|---|---|---|---|---|---|---|---|---|
| Temp (K) | 225 | 230 | 240 | 250 | 260 | 270 | 298 | 400 | 500 |
| $\Delta G_{RXN-1}$ (KJ/mol) | 450.09 | 263.72 | 248.47 | 245.6 | 243.64 | 241.85 | 237.14 | 223.95 | 219.05 |
| $\Delta H_{RXN-1}$ (KJ/mol) | 1804.3 | 412.25 | 303.77 | 293.13 | 290.03 | 288.48 | 285.83 | 242.85 | 243.83 |
| $\Delta S_{RXN-1}$ (KJ/mol K) | 6.02 | 0.65 | 0.23 | 0.19 | 0.18 | 0.17 | 0.16 | 0.05 | 0.05 |
| $T\Delta S_{RXN-1}$ (KJ/mol ) | 1354.2 | 148.53 | 55.29 | 47.53 | 46.39 | 46.63 | 48.69 | 18.89 | 24.77 |
| $E_{thermoneutral}$ (V) | 9.35 | 2.14 | 1.57 | 1.52 | 1.50 | 1.49 | 1.48 | 1.26 | 1.26 |
| $E_{emf}$ (V) | 2.33 | 1.36 | 1.28 | 1.27 | 1.26 | 1.25 | 1.23 | 1.16 | 1.14 |

**For $CO_2$-brine electrolysis**

$$CO_2 + 2H_2O = CH_4 + 2O_2$$

The dH, dS and dG values in table S3 and S4 are calculated from the below expressions[2,3],

$$C_p = 0.039 * T + 25.743 \quad (\text{for} - CO_2) \ (2)$$

$$C_p = 0.0001 * T^2 - 0.038 * T + 36.839 \quad (\text{for} - CH_4) \ (3)$$

**Table S3: Estimation of $CO_2$ (g) thermodynamic properties for $CO_2$-brine electrolysis**

| Temp | dH (kJ mol-1) | dS (kJ mol-1) | dG (kJ mol-1) |
|---|---|---|---|
| 225 | -396.042 | -0.00681 | -394.51 |
| 230 | -395.882 | -0.00610 | -394.479 |
| 235 | -395.721 | -0.00540 | -394.452 |
| 240 | -395.558 | -0.00471 | -394.428 |
| 250 | -395.226 | -0.00334 | -394.39 |
| 260 | -394.886 | -0.00201 | -394.364 |
| 270 | -394.538 | -0.000689 | -394.351 |
| 280 | -394.182 | 0.000605 | -394.351 |

**Table S4: Estimation of $CH_4$ (g) thermodynamic properties for $CO_2$-brine electrolysis**

| Temp | dH (kJ mol-1) | dS (kJ mol-1 k-1) | dG (kJ mol-1) |
|---|---|---|---|
| 225 | -77.354 | -0.090 | -57.011 |
| 230 | -77.188 | -0.09 | -56.561 |
| 235 | -77.024 | -0.089 | -56.115 |
| 240 | -76.858 | -0.088 | -55.672 |
| 250 | -76.525 | -0.087 | -54.797 |
| 260 | -76.189 | -0.086 | -53.936 |
| 270 | -75.85 | -0.084 | -53.087 |
| 280 | -75.506 | -0.083 | -52.25 |

**Table S5: Estimation of Thermoneutral voltage and electromotive force from the thermodynamic properties for $CO_2$-brine electrolysis.**

| Temp (C) | -48 | -43 | -33 | -23 | -13 | -3 | 25 | 127 | 227 |
|---|---|---|---|---|---|---|---|---|---|

| Temp (K) | 225 | 230 | 240 | 250 | 260 | 270 | 298 | 400 | 500 |
|---|---|---|---|---|---|---|---|---|---|
| $\Delta G_{RXN-3}$ (KJ/mol) | 1237.69 | 865.36 | 835.69 | 830.81 | 827.70 | 824.97 | 817.88 | 800.49 | 800.26 |
| $\Delta H_{RXN-3}$ (KJ/mol) | 3927.31 | 1143.20 | 926.24 | 904.97 | 898.76 | 895.64 | 890.29 | 801.30 | 800.52 |
| $\Delta S_{RXN-3}$ (KJ/mol K) | 11.95 | 1.21 | 0.37 | 0.29 | 0.27 | 0.26 | 0.24 | 0.0020 | 0.0005 |
| $T\Delta S_{RXN-3}$ (KJ/mol ) | 2689.62 | 277.84 | 90.54 | 74.16 | 71.06 | 70.66 | 72.41 | 0.81 | 0.25 |
| $E_{thermoneutral}$ (V) | 5.09 | 1.48 | 1.19 | 1.17 | 1.16 | 1.16 | 1.15 | 1.04 | 1.04 |
| $E_{emf}$ (V) | 1.60 | 1.12 | 1.08 | 1.08 | 1.07 | 1.07 | 1.06 | 1.05 | 1.04 |

## $CO_2$-brine electrolysis operating at different conditions

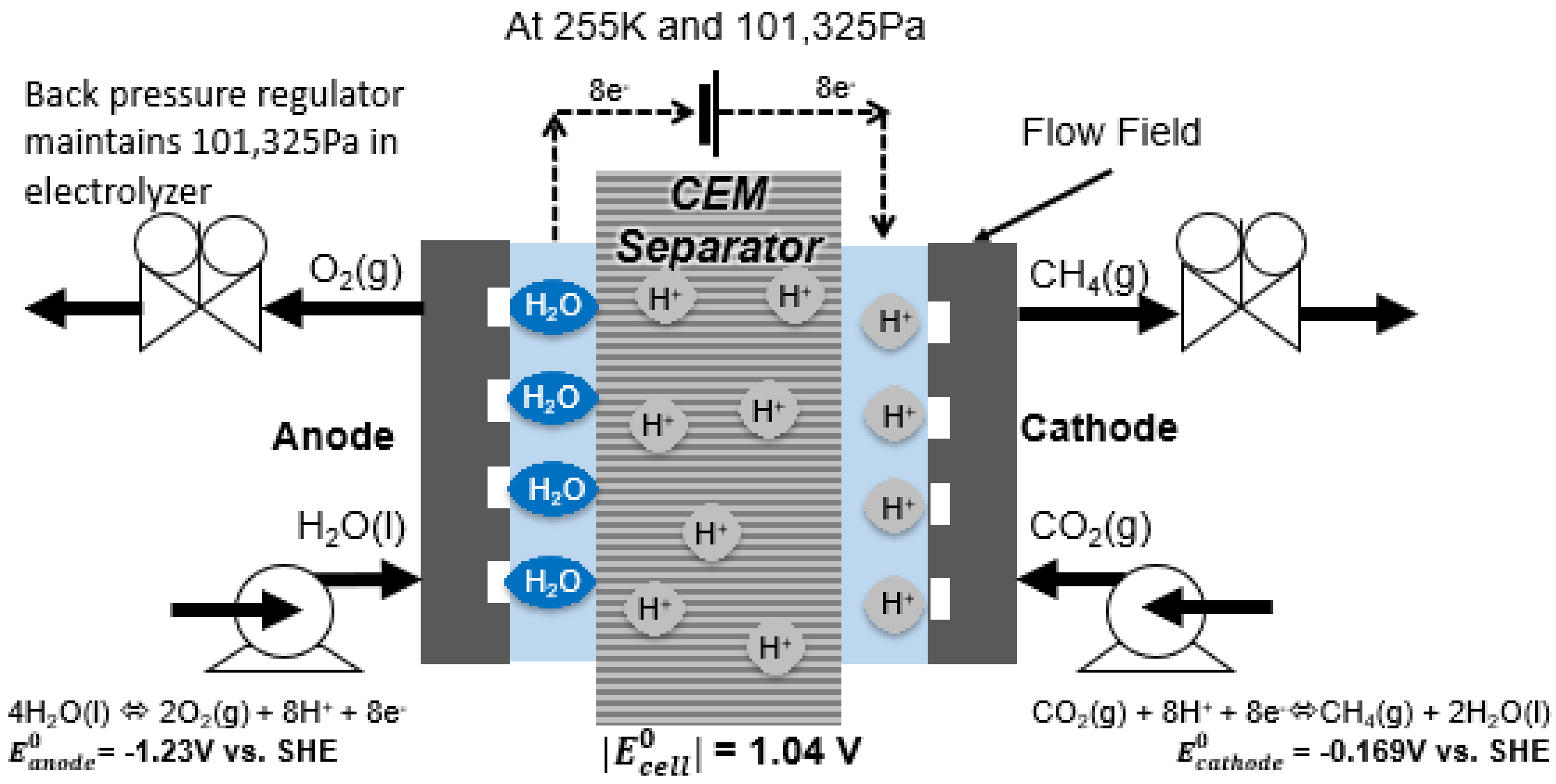

**Figure S1**: $CO_2$ – brine electrolyser operating at Earth atmospheric pressure and Martian temperature.

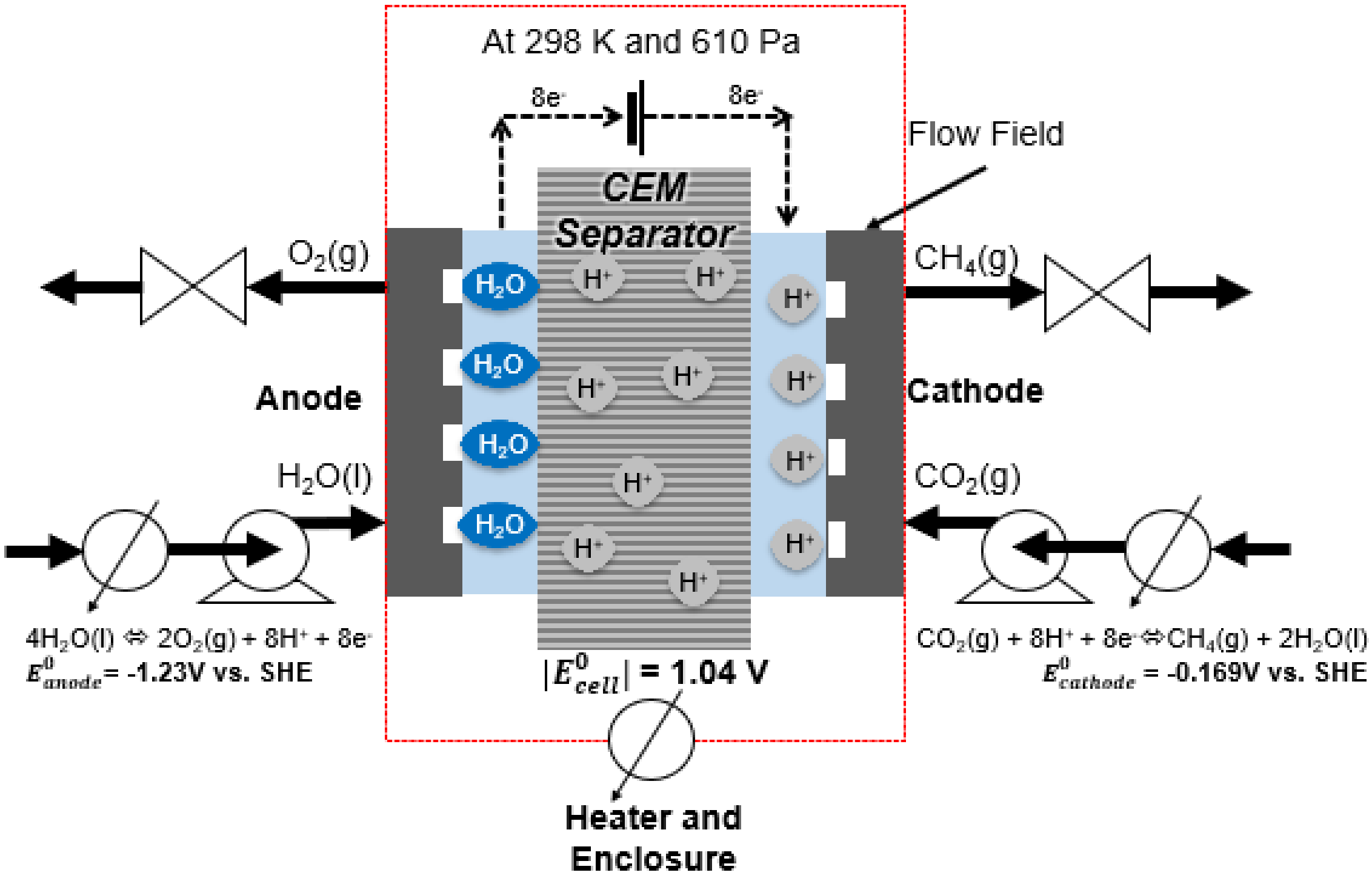

**Figure S2**: $CO_2$ – brine electrolyser operating at Martian atmospheric pressure and standard Earth temperature.

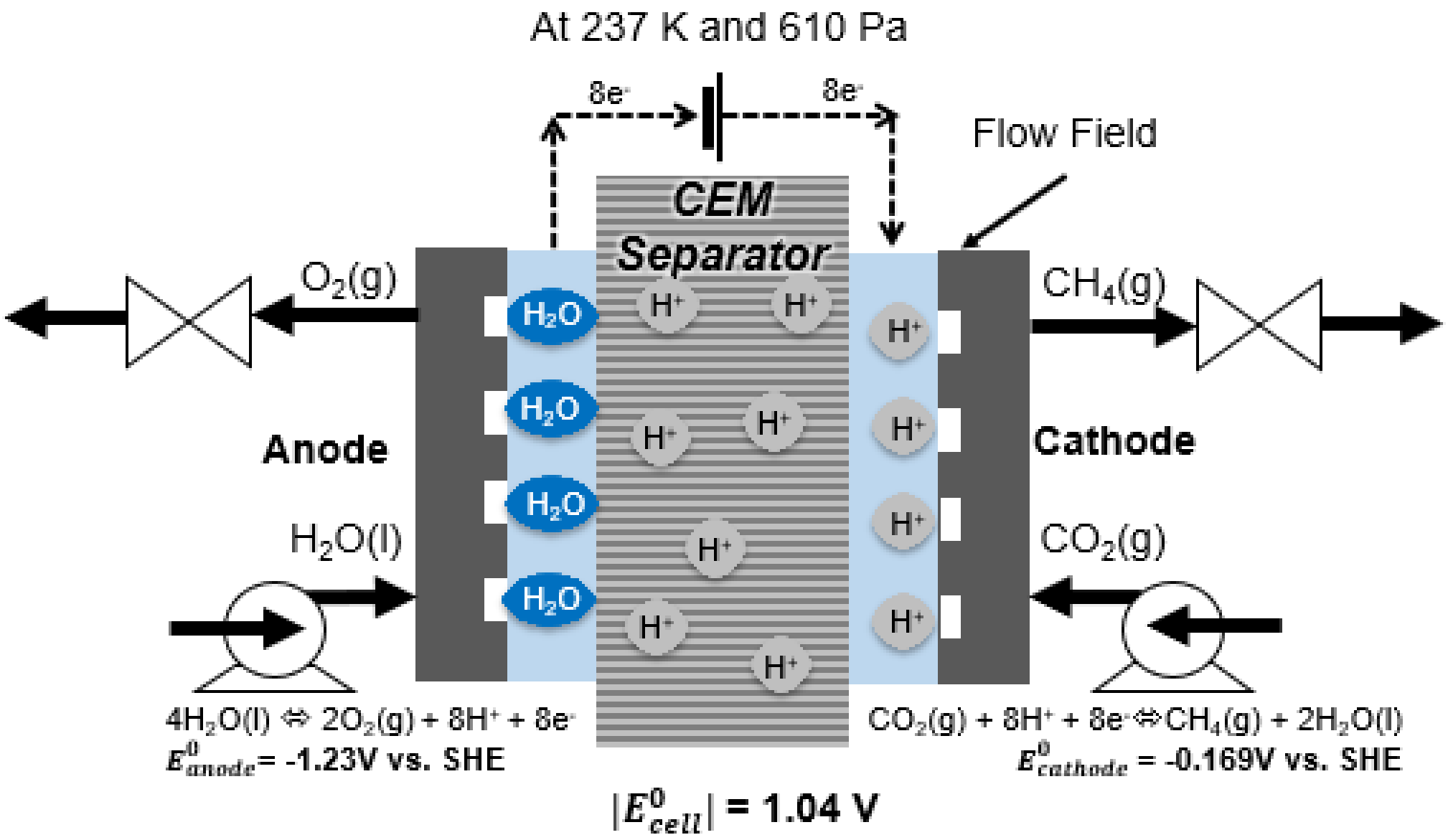

**Figure S3**: $CO_2$ – brine electrolyser operating at Martian atmospheric pressure and temperature.

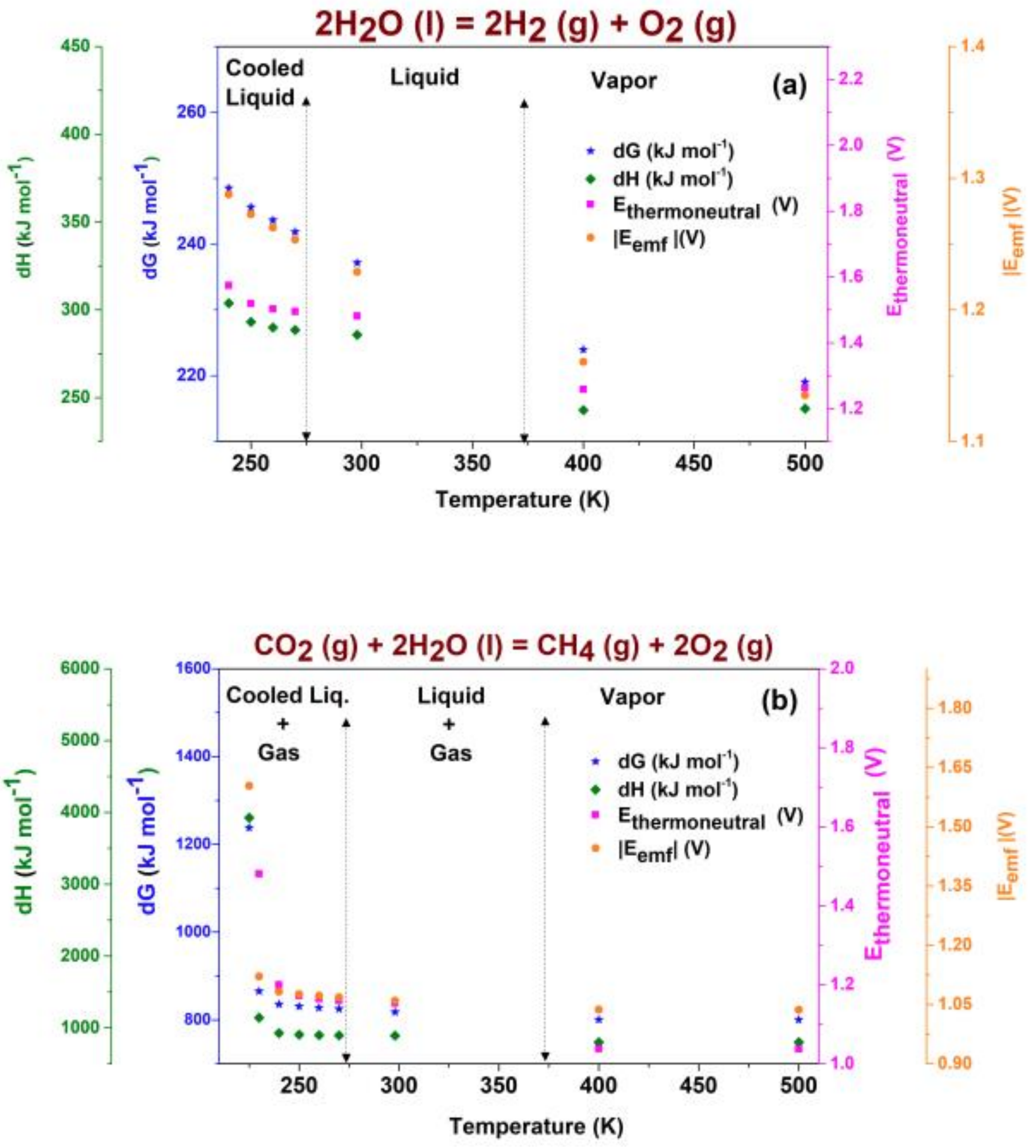

**Figure S4**: Thermodynamic properties for (a) brine electrolysis and (b) $CO_2$ - brine electrolysis between 230K and 500k.